\documentstyle[12pt,epsfig,preprint]{aastex}

\slugcomment{To appear in 
}

\shorttitle{XMM Observation of N LMC 1995}
\shortauthors{Orio et al.}

\begin{document}


\title{A XMM-Newton observation of Nova LMC 1995, a bright supersoft X-ray source}


\author{Marina Orio}
\affil{INAF - Turin Astronomical Observatory, Strada Osservatorio 20, I-10025 Torino, Italy\\
and\\
Department of Astronomy, University of Wisconsin, 475 N. Charter Str., 53706 Madison WI,  USA}
\email{orio@cow.physics.wisc.edu}

\author{Wouter Hartmann}
\affil{SRON Laboratory for Space Research, Utrecht, The Netherlands}

\author{Martin Still}
\affil{NASA Goddard Space Flight Center, Greenbelt MD 20771, USA,
 and  Universities Space Research Association}
\and

\author{Jochen Greiner}{
\affil{Max Planck Institute for Extraterrestrial Physics, Garching bei
M\"unchen, FRG}


\begin{abstract}

Nova   LMC  1995, previously detected during
1995-1998 with  {\sl ROSAT},  was observed  again
as a luminous supersoft X-ray source with {\sl XMM--Newton} 
in December  of 2000. 
 This nova offers the possibility to observe the spectrum
of a hot white dwarf, burning hydrogen
in a shell and not obscured by a wind or
 by nebular emission like in other supersoft X-ray sources.
 Notwithstanding uncertainties
in the calibration of the {\sl EPIC} instruments
at energy E$<$0.5 keV, using atmospheric models in Non Local
 Thermonuclear Equilibrium we derived  
an effective temperature in the range  400,000-450,000 K,
 a bolometric luminosity  L$_{\rm bol} \simeq 2.3 \times 10^{37}$ erg s$^{-1}$,  
 and we verified that the abundance of carbon 
is not significantly enhanced in the X-rays emitting shell.
 The {\sl RGS} grating spectra do not show  emission lines
(originated in a nebula or a wind) observed for some other supersoft X-ray sources. 
The crowded atmospheric absorption lines of the white
 dwarf cannot be not resolved.  There is no hard component (expected
from a wind, a surrounding nebula  or an accretion disk), with no counts above
the background at E$>$0.6 keV,
 and an upper limit F$_{\rm x,hard} = 10 ^{-14}$ erg s$^{-1}$ cm$^{-2}$
 to the X-ray flux above this energy. The background corrected
count rate measured by the {\sl EPIC} instruments was variable
on time scales of minutes and hours, but
 without the flares or sudden obscuration
observed for other novae. The power spectrum shows a peak
at 5.25 hours, possibly due to a modulation with the orbital period.
We also briefly discuss the scenarios in which this
 nova may become a type Ia supernova progenitor.

\end{abstract}

\keywords{stars: novae, cataclysmic variables - X-rays: stars}

\section{Introduction}

Nova LMC 1995 was discovered in outburst in the Large Magellanic Cloud at the 
beginning of March 1995 (Liller 1995). It 
reached V$\leq$10.7  at maximum brightness (above  average for a  LMC nova), and
the expansion  velocity  was in the range 800--1500 km  s$^{-1}$
(Della Valle et al. 1995).   The rise to maximum  took at least a  few
days. The decay by one  magnitude in almost  3 days indicated a
moderately fast or fast nova (see  Liller 1995, Gilmore 1995, Christie
1995), but only sparse observations were done and the subsequent optical
lightcurve is not known. Luminous, supersoft X-ray emission
was discovered with the X-ray satellite {\sl ROSAT} by two of
 us (Orio \& Greiner 1999) three years after the outburst. Even
if post-novae white dwarfs are expected to appear as a supersoft
X-ray source for some time, for most of them this phase seems to
 be short lived (see Orio et al., 2001). N LMC
1995 was an interesting exception and deserved to be further monitored. 

  Classical novae are
cataclysmic variables, that is close binary systems in which a white dwarf
 accretes matter from a companion filling its Roche lobe. Novae
undergo outbursts of amplitude $\Delta$m=8-15 mag in the optical range; 
the total energy emitted is 10$^{44}$-10$^{46}$ erg
 (a nova is the third most energetic
phenomenon in a galaxy after gamma ray
bursters and supernovae). The outbursts are thought to be triggered
on the white dwarf  by a 
thermonuclear runaway in the hydrogen burning shell at the bottom of the 
accreted layer. A 
radiation driven wind follows, depleting all or part of the accreted 
envelope (see Kovetz 1998, Starrfield 1999).
 Residual hydrogen burning in a shell on the white dwarf occurs unless
 all the envelope is ejected after the outburst, while the atmosphere 
shrinks 
and the effective temperature increases. The post-nova appears as a 
very hot blackbody-like object with  effective temperatures 2.5 $\times$ 10$^5$-10$^6$ K 
(see Prialnik 1986), and luminosity in the range 10$^{36}$-10$^{38}$
erg s$^{-1}$.  The duration of the supersoft X-ray phase  
is probably directly proportional to the leftover envelope mass.
If some of the accreted mass envelope
is retained after each outburst, the white dwarf mass increases towards the
Chandrasekhar mass after a large number of outbursts in the same system, 
eventually leading to a type Ia supernova event or to the formation of a 
neutron star by accretion induced collapse.
 The supersoft X-ray luminosity is the only clear 
 indication of how long the hydrogen rich fuel lasts.  X-ray
observations up to now indicate that most novae do not
 keep a significant amount of mass after each outburst
(see Krautter 2002, Orio et al. 2001). 
Only one Galactic nova has been observed as a supersoft
X-ray source after more than 2 years: the {\sl ROSAT PSPC} detected 
  GQ Mus ( N Muscae 1983) 
9 years after the outburst. The X-ray flux however
decayed towards ``turn-off'' in the following year 
(\"Ogelman et al. 1993, Shanley et al. 1995). 

Nova LMC 1995 was  the only LMC nova  detected as a luminous supersoft
X-ray source in   the Magellanic Clouds  in  repeated pointings and  a
{\sl ROSAT} survey of  the two galaxies (see Orio \& Greiner,
1999, Orio et al., 2001).   It was observed already 
before the eruption, but it
was only detected with the {\sl ROSAT HRI} for the first time 9 months after
the outburst. 
Orio \&  Greiner  (1999) showed   that  the data 
taken in February of 1998 could   be  fitted with an
atmospheric  model of a  $\simeq$ 1.2 M$_\odot$  white dwarf with an
temperature T$_{\rm eff} \simeq$ 345,000 K.  The X-ray flux increased
in the first 3 post--outburst years. The interpretation is that 
the  atmosphere kept on shrinking,  as  its temperature increased.
  In Orio \& Greiner (1998), we
suggested that  N LMC 1995  may  be the prototype  of  a rare class of
novae that are bound to reach the Chandrasekhar mass, becoming type Ia
SN  or even undergoing accretion induced  collapse. Therefore, it appeared 
worthwhile to follow the subsequent evolution of N LMC 1995. 
Nova white  dwarfs that turn   into  supersoft X-ray sources also  offer  
the only possibility to determine  the white dwarf parameters
using white dwarf atmospheric models.

In this paper we will describe mainly observations done
with {\sl XMM-Newton} at the end of the year 2000. 
 The purpose of the observation was not only to measure
the length of  the supersoft X-ray phase, but also  to obtain the  physical
parameters of the  white  dwarf from the  {\sl EPIC}  and {\sl RGS} spectra.

\section{The last {\sl ROSAT PSPC} observation}

In  December  of 1998, before the   {\sl PSPC} was turned  off, N  LMC 1995 was
observed one last time. The high-voltage drop-out at this stage caused
calibration problems. We measured  a  background corrected count  rate
0.030$\pm$0.004 counts  s$^{-1}$. This value is 
lower by a factor  of 2 compared to the
one of February 1998, however the calibration at the end of the PSPC
life had become 
uncertain and the usable exposure was short (only 2408 s),
so we could not conclude that the flux
 had decreased.  The source  was certainly
still  very luminous in the supersoft X-ray range.  This
observation prompted us to propose a new one to be done with {\sl XMM-Newton}.

\section{Observations with {\sl XMM-Newton}}

N  LMC 1995 was observed with  {\sl XMM-Newton} two years later, on December
19 2000, almost 6 years after the outburst.
A description of the mission can be found in Jansen et al. (2001).
The satellite carries an optical telescope, which was shut off
due to the presence of a bright star in the field, and three
X-ray telescopes with five
X-ray detectors, which were all used: the European Photon Imaging
Camera ({\sl EPIC}) {\sl pn}
(Str\"uder et al. 2001), two  {\sl EPIC MOS} (Turner et al. 2001),
 and two Reflection Grating Spectrometers (den Herder et al.
2001). The observation lasted for 46400  seconds with  {\sl EPIC MOS},
and 46280 seconds with {\sl EPIC pn}.
 The data were reduced with the {\sl ESA XMM Science Analysis System
(SAS)} software, version 5.3.3., using the latest calibration
files available in June of 2002. The {\sl EPIC} data were taken in
the prime mode, with full window, and the thin filter.
  
We still detected the luminous supersoft X-ray source.
The average count rates
 measured with the different instruments are shown in Table 1, and
the observed {\sl EPIC} and {\sl RGS} spectra are shown in Fig. 1 and 2,
respectively. The RGS energy range is 0.35-2.5 keV, and for N LMC 1995
there is no significant S/N above $\simeq$0.48 keV. The
 corresponding useful wavelength range is 26-35 \AA. We indicated wavelength
instead of energy
in Fig. 2 to facilitate the comparison with Paerels et al. (2001), Bearda et al. (2001), 
and Burwitz et al. (2002). 

   We notice that this spectrum is peaked at lower
energy than most supersoft X-ray sources, and can be compared
only with the spectrum of nova V382 Vel observed with BeppoSAX
10 months after the outburst (Orio et al.
2002). In the case of V382 Vel, however, a low luminosity, hard component
of nebular origin was also present, that is absent for N LMC 1995.

\subsection{Time variability}

The {\sl EPIC} and {\sl MOS-1} background corrected lightcurves are
 shown in Fig. 3. The 
lightcurve background was extracted from a ring around the source and
normalized to the extraction area used for the source. 
We rule out large flares (observed for V1494 Aql, see Starrfield et al. 2001,
or Drake et al. 2002), or a sudden obscuration
(observed for V382 Vel, see Orio et al. 2002). However,
we found that the count rate varied by more than  a factor of five 
during the $\simeq$46 ksec of observation. There are 
 irregular variations on time scales of few minutes 
in the first portion of the light curve as
well as a modulation with a longer time scale. The power spectrum
 shows the highest peak at 18900$\pm$100 seconds
 (5.25 hours), which is  probably related to the orbital period 
 (a study of the optical lightcurve is in preparation, Orio \& Lipkin 2003).
The modulation, which must be dependent on the inclination, is of
 larger amplitude than the X-ray eclipses observed for Cal 87
(Schmidtke et al. 1993), and possibly for GQ Mus (see Kahabka 
1996; note that this result is uncertain because the orbital period
of the system is close to the orbital period of the {\sl ROSAT} spacecraft),
 but it is comparable to the one in  the X-rays lightcurve of SMC 13
(Kahabka 1996).  Other peaks found in the power spectrum
are only aliases of the exposure time, and we did 
not detect the pulsations observed for
 V1494 Aql (Drake et al. 2002).

\subsection{Comments on the {\sl RGS} spectra}

In Fig. 2 we show the portions of {\sl RGS} spectra in the 25--35 \AA \
range, for which the signal to noise ratio (S/N) 
is not too low for fitting model atmospheres.
The wavelengths of the most important absorption lines in a thermal
 plasma at T$\simeq$407,000, obtained
with the model fits (see Table 2),  are labelled in the figure.
At this temperature the he CNO elements must be
 in their H- and He-like charge states. 
 Unfortunately, not only is the S/N ratio very modest, but
the {\sl RGS} does not have the spectral resolution
(0.05  \AA \ or less) needed to resolve most of the
intricate line spectrum of an extremely hot white dwarf atmosphere. 

 However, these spectra are useful, also because 
 the relative calibration of the two {\sl RGS} is better determined
than the {\sl EPIC} calibration at the moment. For {\sl RGS-1} and {\sl RGS-2} 
the uncertainties in the relative calibration across the
 instruments are not greater than 7\% and 5\% respectively
in the 20-28 \AA \ range, and less than 12\% and 10\% in the 28-36 \AA \ range. 
 Moreover the {\sl RGS} spectra, as it can be seen in  in Fig. 2, also allow us to rule out the 
presence of strong, narrow lines {\it in emission} 
observed in the {\sl Chandra} observation of Nova V382 Vel (due to 
 the surrounding nebula, see Burwitz et al. 2002)
and in the {\sl Chandra} and {\sl XMM} exposures
of the supersoft X-ray source MR Vel (due to a wind, see Bearda et al. 2002
 and Motch et al. 2002).  We note that there is only one LMC supersoft source
which is significantly brighter than N LMC 1995
 in this spectral range, Cal 83, for which {\sl RGS}
count rate is $\approx$10 times higher, but this is partially due to
the source being harder (Paerels et al., 2001).  
A qualitative comparison of the {\sl RGS} spectra of N LMC 1995
 and Cal 83 shows some similarities but also  a large difference.
The continuum for N LMC 1995 is fitted with atmospheric models 
with line opacities (see next
section), but these models do not fit the continuum  
 adequately for Cal 83 (Paerels et al. 2001).  
Not resolving the line spectra, detailed comparisons are not possible. 
Qualitatively, however, the {\sl RGS} spectrum in the 
25-35 \AA ~ region appears quite different for Cal 83 and N LMC 1995.

\subsection{Fitting white dwarfs atmospheric models to the observed spectra}

The continuum
of the {\sl EPIC} spectra detected with the {\sl pn} and the two {\sl MOS}
can be fitted with atmospheric models, to derive the white dwarf
parameters and its chemical composition. The {\sl EPIC pn}
absolute calibration in the lowest energy range is still being improved. 
Before launch, the uncertainty in the
 relative calibration of {\sl EPIC-pn} and {\it EPIC MOS} was
only 1\% in the 0.2-0.5 keV range. However, in
the {\sl MOS} the X-rays absorbed near the
surface layers can loose a large fraction of their charge before collection,
in a process which
is not related to charge transfer inefficiency across the CCD
(Breitfellner 2002, private communication). This effect is
one of the dominating factors for the calibration and a large
 number of counts below 0.3 keV for sources
 with column density N(H)$\geq 10^{20}$ cm$^{-2}$ is due to scattered  
photons from higher energies. Extensive ground calibrations were used to construct
 a pre-flight response
matrix, but after the launch the surface fractional
charge loss at low energies was greater than observed on the
ground. It also seems that the surface loss effect is not constant with epoch
and observations of sky calibration
sources after Orbit 300 reveal an excess of counts in the soft band.
In a future release of the {\it SAS} software, the {\sl MOS} calibration
will be adjusted to incorporate the epoch dependent surface charge
 loss (Breitfellner 2002, private communication), but it is not
available yet at present.   

We used the {\sl XSPEC} software package (Arnaud 1996). In the first
 two years after the outburst all novae seem to emit
 hard X-rays, but 6 years after the outburst, for this nova the 
3 $\sigma$ upper limit to the X-ray flux in
the range 0.6-10 keV is of $\simeq$10$^{-14}$ erg cm$^{-2}$ s$^{-1}$.
 The X-ray luminosity in this range is therefore
L$_{\rm x,hard} \leq 3 \times 10^{33}$ erg s$^{-1}$.
 We remind that N Pup 1991 was still emitting a higher hard X-ray
luminosity then this upper limit 16 months after the outburst (Orio et al. 1996).  
 
For the emission below 0.6 keV, we find that the black-body
in {\sl XSPEC} is not adequate to explain the shape
of the observed continuum.  We used Local
Thermodynamical  Equilibrium (LTE)   models  
of Heise et al. (1994), Non-LTE (NLTE) models 
developed   by  Hartmann \& Heise
(1997) and Hartmann et al.
(1999), and also other   metal enhanced NLTE atmospheric models  
developed by one of us (W.H.) for this  project, and we implemented these
models  in {\sl XSPEC}.
 Optically thick model spectra that only involve continuum opacities 
usually do not represent the observed spectrum correctly.
We also included the effect of many line opacities (line blanketing), 
that change the temperature structure of the model atmosphere, 
therefore the ionization balance and eventually the shape of the model spectrum.

In our case, like for the observations  done by other authors
(e.g. Hartmann \& Heise 1997), a better fit is 
definitely obtained with the NLTE models than with the LTE ones. 
  The NLTE model atmospheres are calculated using the computer code TLUSTY 
(Hubeny 1988, Hubeny \& Lanz 1995).
 This code uses the complete linearization technique to solve the
coupled set of radiative transfer, radiative equilibrium and statistical equilibrium equations.
Convergence is achieved when the relative changes of the temperature,
 total number density and electron density are smaller than 10$^{-3}$.
 For a detailed
description of the computer code TLUSTY we refer to Hubeny
(1988) and Hubeny  \& Lanz (1995).

Lanz \& Hubeny (1995) show that metal line opacities
for carbon and iron have to be
 taken along in NLTE model atmosphere calculations. They conclude that line blanketing by trace
elements with abundances above  solar influences the atmospheric structure of hot,
 metal-rich white dwarfs. Rauch (1996) calculated line-blanketed NLTE model
atmospheres, to show that light metal opacities drastically decrease the flux
 levels of hot stars.
          We restrict ourselves to a limited number of ionization stages and atomic levels.
 The spectrum of hot, high-gravity atmospheres is often dominated
 by the lowest levels
of one or two ionization stages of a particular element. Therefore, we selected the
 ionization stages that we expected to be most dominant in the 
range of parameters of interest.
Table 2 shows the ionization stages included for the 
ions of the elements considered (H, He, C, N, O, Ne, Mg, Si, S, Ar,
Ca, and Fe). The atomic data are from the 
Opacity Project Database (Cunto et al., 1993).  
 We refer to Hartmann et al. (1999) for the detailed treatment
of continuum opacities.

To fit the N LMC 1995 {\sl XMM} spectra,
we tested a grid of models for log(g)=8, 8.5, 9, interpolating between 
 temperature steps of 50,000 K, with the following
sets of abundances:

a) cosmic abundances (Anders \& Grevesse 1989);

 b) ``LMC-like'' abundances, (from Dennefeld, 1989,
roughly 0.25 times the cosmic values);

 c) C, N and O enhanced by a factor of 10 with respect to ``cosmic'';

d) Ne, O and Mg enhanced by a factor of 10; 

e) enhanced He (H/He=0.5 in number abundance)  and enhanced
C/N ratio (a model atmosphere developed for U Sco, see Kahabka et al. 1999).

We adopted the strategy of fitting the data of the
 different instruments separately at first, then we combined 
the {\sl EPIC pn} and {\sl MOS} spectra. Finally, we fitted the spectra
of all 5 instruments together. We examined and fitted the 
the {\sl RGS} spectra only in the range 0.35-0.48 keV (26--35 \AA), where
 the signal to noise ratio is large,
but still acceptable, because the rest of the spectrum
 is too noisy. In Table 3 we report all the
results of spectral fits that yielded $\chi^2$/d.o.f. $\leq$ 1.5. No models
fit the {\sl MOS} spectra with $\chi^2$/d.o.f. $\leq$ 1.6 if we include
 the range 0.2-0.3 keV. We attributed this 
to a difference in the relative calibration of the {\sl MOS} and
{\it pn} at this energy, probably
 due to the surface layers
absorption mentioned above. Since fitting the {\sl pn} spectrum
 at 0.2-1 keV decreases $\chi^2$/d.o.f. a little, but the qualitative results 
agree quite well with the results in the 0.3-1 keV range (see Table 3), we
tried including the 0.2-0.3 keV range for the {\sl pn}. 
When we did, we obtained a higher value of the 
equivalent column of neutral hydrogen, N(H). 
We set some limits inside which we accept that the N(H) parameter may vary:
N(H)=0.7--1.8 $\times$ 10$^{21}$cm$^{-2}$, 
from the Galactic absorption to the LMC, 
to this value plus the intrinsic absorption in    the LMC,
 evaluated  with  Points' (2002) ATCA-Parkes
H  I maps (which show that  the N(H) column all  the way to the
back of the LMC in the direction of the nova is 1.1 $\times$ 10$^{21}$
cm$^{-2}$.) 

The best fit T$_{\rm eff}$ and N(H) with their 2 $\sigma$ uncertainties
are given in Table 3, for models
which yield $\chi^2$/d.o.f.$\leq$1.5.  The value of T$_{\rm eff}$ varies
in a narrow range in all the models and the uncertainties on
the absorbed flux typically do not exceed 25\%.
The last two columns of Table 3 indicate the parameters that we used
 to determine what model fits the spectrum. One parameter is
$\chi^2$/d.o.f. (last column). The column before the last lists
also the other important parameter that we had to consider: the
normalization factor K = (R$_{\rm WD}$/d)$^2$, where d is the distance to
 the LMC (here we adopted 51 kpc, but a distance of 55 kpc, adopted by
other authors, does not change our general conclusions).
 Deriving the radius R$_{\rm WD}$ from K, we obtain the white
dwarf mass M$_{\rm WD}$ knowing the effective gravity g
(which is a model characteristic) through   M$_{\rm WD}$ = g R$_{\rm WD}^2$/G .
The normalization factor K must indicate  a mass 
M$_{\rm WD}$ in the range 0.6--1.4 M$_\odot$.
All models with a value of ``K'' that yields  M$_{\rm WD}$ outside this range 
are not consistent with the LMC distance. The
upper limit is the Chandrasekhar mass, the lower one the minimum mass to
obtain a sufficient  density in the shell to allow
thermonuclear burning of hydrogen.
At very high effective temperature, for the hydrogen burning shell to
 radiate all the energy it produces, R$_{\rm WD}$ must be  
bloated to some extent.  It may be quite larger than the Chandrasekhar radius
(e.g. Prialnik 1986, MacDonald et al. 1985). 
We note that the value of K increases dramatically with N(H), but keeping 
constant N(H),  K increases also with the effective gravity. N(H) and K,
unlike the temperature and the absorbed flux, vary by a large
extent for the different models.

All the atmospheric models  with enhanced carbon, either
in LTE or NLTE (specifically
models ``c'' or ``CNO enhanced'' and ``e'' or U Sco-like), are
not included in Table 3 except in one case for the {\sl RGS}, 
because they cannot  fit  the {\sl EPIC} data. This
 is mainly due to  a too deep  absorption edge   of C VI   at 0.49  keV.  
Therefore, we find that carbon is not enhanced. The continuum predicted by model ``e'' differs
mostly from the observed continuum, and this is due mainly to the 
 enhanced C/N ratio (more than to the helium enhancement). 

Regardless of the value of the ``K'' normalization factor,
  the best  fit  to the {\sl EPIC pn} spectra  and to
all the instruments simultaneously, is  obtained  with the NLTE  model  atmosphere ``d'' 
 with log(g)=9, at  T$_{\rm eff} \simeq$ 450,000 K,
and N(H) = 1.6  $\times 10^{21}$ cm$^{-2}$.  This model fits the data  
better due to a deeper edge of Mg X at 0.347 keV, 
and one of Mg IX at 0.23 keV, which affect the continuum slope
at low energy. Thus is it is possible that 
magnesium is enhanced, but the fit had to be discarded due to 
the value of the constant K, which is definitely too high.  
This model indicates a high column density N(H) and so  high a flux that the 
white dwarf mass would exceed the Chandrasekhar mass by a great amount. The same is true even
for the second best fit, the same model with log(g)=8.5.
Model ``d'' with log(g)=8 does not fit the {\sl EPIC} spectra adequately.
  Even if the next good model, ``b'' with log(g)=8,
 fits the spectrum a little less well (especially due to 
a poorer result if we include the spectral range
0.2-0.3 keV for the {\sl EPIC pn}) the constant K
is finally  acceptable.  The value of K for the best fit to all
the instruments, including the 0.2-0.3  keV range
of the {\sl pn},  indicates M$_{\rm WD}$ = 0.905 M$_\odot$.
The unabsorbed flux in the 0.2-1.0 keV band is F$_{\rm x} = 8 \times
10^{-13}$ erg cm$^{-2}$ s$^{-1}$, and
the bolometric luminosity is L$_{\rm bol} = 2.3 \times 10^{37}$ erg s$^{-1}$
at a distance  51 kpc.
 
We note that all the fits with the atmospheric models, in LTE and NLTE,
for all instruments, indicate 
 3.9 $\times 10^5 \leq$ T$_{\rm eff}\leq 4.7 \times 10^5$ K,
but the {\sl RGS} spectra
are  always best fitted with T$_{\rm eff} > 4 \times 10^5$ K. 
Comparing with the results for the {\sl PSPC} data in Orio
\& Greiner (1999), this clearly indicates that  
the WD atmospheric temperature has remained at least constant 
(within the 2$\sigma$ confidence level) or,
most probably, has increased.

A post-nova white dwarf which is still burning hydrogen in a shell
is expected to evolve at constant bolometric luminosity. 
 While the radius shrinks and the atmosphere
becomes hotter, 
the flux in the 0.2-1 keV range is supposed to remain constant. 
 An unabsorbed bolometric flux F$_{\rm bol}$ of   order of  $\simeq$5 $\times
10^{-10}$ erg cm$^{-2}$ s$^{-2}$,   was consistent     
 with the fit  of Orio \& Greiner (1999; note that  the fit was done
with models at fixed bolometric luminosity, unlike
the models used here for which the normalization
constant is a parameter). This
is an order of magnitude larger than F$_{\rm bol}$ we derived
here, but we have to stress that the 1$\sigma$ uncertainty
 in the {\sl ROSAT} spectral fit parameters translates in an uncertainty
of more than one order of magnitude  in the  flux, so we
ignore whether  the peak temperature was reached {\it between}
the {\sl ROSAT} and the {\sl XMM-Newton}  
observations and whether the flux at the end of  2000 had started to decrease.
The models used in Orio \& Greiner (1999) are LTE models
of MacDonald \& Vennes (1991). The 
white dwarf  in the model had  log(g)=7 and
R$_{\rm WD}$=2.5 $\times 10^{10}$ cm. The white dwarf radius is only
R$_{\rm WD}$=1.09 $\times 10^9$ cm  in our ``b'' model with log(g)=8.
Despite the different models used, clearly
the evolutionary picture  that emerges is that
of a shrinking atmospheric radius in the period from the
beginning of 1998 and the end of 2000, and it is consistent with the 
theoretical prediction.
\section{Conclusions}

The   spectrum of the post-nova  supersoft  X-ray sources in the first
year or two after  the outburst can be very  complex.  The X-ray  flux
in other novae around one year after the outburst has been observed to vary 
dramatically on short time  scales, 
 and   this is  not  well understood yet   (Orio et
al. 2002, Drake et  al. 2002). The central  source may be hidden by
another source of X rays, thought to be the ejected
 nebula, with emission lines in the supersoft X-ray range due to transitions of
highly ionized elements. These narrow emission lines were resolved
only in one case  with the {\sl Chandra LETG}
(Burwitz et al. 2002, see also the discussion of 
 Orio et al.  2002).  Since the BeppoSAX LECS  and the {\sl ROSAT  PSPC}
could  not resolve super-imposed nebular  narrow emission lines from the central source
spectrum, temperature   and  effective gravity    determined from the
observations of V1974  Cyg (Balman et al. 1998)
and  of V382    Vel (Orio et   al.   2002) should be  considered  very
uncertain. N LMC 1995,  at the late 
post-outburst stage at which it was observed, appears as a
``simpler'' source, without nebular emission. 
Grating observations of another supersoft X-ray source have also shown
narrow  emission lines, most likely due to
an ongoing stellar wind that emits X-rays and shields  the white dwarf continuum,
(Bearda et al.  2002).  These lines are also absent for Nova LMC 1995.
Moreover, despite some variability (probably connected with
the orbital period), sudden flares or obscurations are not observed.
 Thus this nova  offers a rare, perhaps unique 
 opportunity to observe the ``naked'' atmosphere
of a supersoft X-ray nova remnant and determine the white
 dwarf parameters. We consider it a sort of
 ``Rosetta stone'' of hydrogen burning, post-outburst novae.  
 Once the {\sl EPIC} calibration is refined, and 
 the epoch-dependent  {\sl MOS} calibration is made
available, it will be meaningful to try and fit the spectra
with a finer grid for log(g).  The important point is that,
if we determine log(g) more accurately, we also obtain a
precise  estimate of M$_{\rm WD}$.
We remind that we have not yet included detailed NLTE models with log(g)=7.5,
which still have to developed, but since there
is a clear trend of K with log(g), we foresee that for LMC abundances we 
would obtain too low a mass. On the other hand, models with higher abundances,
with the present instrument calibration
do not fit the data with a reasonable value of K for 
{\it any} value of log(g). 

A different, but model-dependent way in which we may 
derive the white dwarf mass, will be to 
measure the atmospheric temperature reached at maximum. 
According to all models (e.g. MacDonald
 et al. 1985), this temperature it is critically dependent on
the white dwarf mass. For this reason,
we have required further monitoring with {\sl XMM-Newton}, and it has been
scheduled for 2003.  
There is also an inverse dependence of the white dwarf mass
on the time to reach the peak temperature.  According to the relationships given
by MacDonald et al. (1985),
 the  range of masses for which the peak effective
temperature is expected to exceed T$_{\rm eff} \geq$407,000 K
(yielded by our most likely model ``b''), 
yet the time during which this temperature increases is
$\geq$ 6 years, is  0.74 M$_\odot \leq$ M$_{\rm WD} \leq$ 1.21 M$_\odot$,
in agreement with the best estimate M $\simeq$ 0.9  M$_\odot$ derived
from the model atmosphere. MacDonald et al. (1985)
predict that a white dwarf of 0.9 M$_\odot$ burns hydrogen in a shell
 for almost 39 years, exceeding a maximum temperature of 550,000 K,
which will be reached slowly while the atmosphere shrinks.   
The new observations, done again with {\sl XMM}, and the
improved {\sl EPIC} calibration, should allow us to
verify this theoretical prediction 
 by measuring the white dwarf mass in two independent ways:
with model atmospheres, and deriving it from the peak temperature and
turn off time.  This will be an important test for the models.

We tested atmospheric models with different abundances. 
If there is significant mixing with the white dwarf material, slowly 
``eroded'' during the secular evolution, 
the white dwarf (which presumably  is a CO or
a Ne-O-Mg white dwarf) may be burning some of its own material
(not accreted from the companion), and be decreasing rather than increasing in mass.
With the present status
of instrument calibration, we were only able to determine that carbon is 
not significantly enhanced. If there is mixing with the outer 
layers   of a CO white dwarf, both these two elements
would be enhanced, but the carbon abundance is the one that
mostly produces a change in spectral shape in the energy
range in which N LMC 1995 emits copious flux. This change
has not occurred.  The spectral fits and the
distance constraints seem to even favor a model with 
depleted abundances, but an improved instrument  calibration will be
necessary to better assess the sophisticated differences
in the models and perhaps find out whether the   burning material may be accreted 
from the companion.  Since shell hydrogen burning has continued immediately
after the outburst, if the hydrogen rich material is not significantly 
 enhanced in heavy elements it cannot have its origin in the WD itself, 
 but it must be accreted material, still  retained after the eruption.

This is important because,
if the white dwarf mass grows, after repeated nova outbursts 
it might reach the Chandrasekhar mass and explode as type Ia SN.
However, we have no indications that M$_{\rm WD}$ in N LMC 1995 is
 anywhere near the Chandrasekhar value yet. If M$_{\rm WD} << 1.4$ M$_\odot$, 
it could take longer than the Hubble time. 
 Another  scenario in which a nova systems eventually undergoes 
a type Ia SN explosion  is a ``sub-Chandrasekhar mass'' model 
(see Fujimoto \& Sugimoto 1982) in which,  
after many nova outbursts,   the supernova explosion 
is triggered by a helium flash. The helium flash is made possible by 
significant accumulation of a helium buffer during previous 
hydrogen burning in the nova secular evolution.
In this work we could not prove, nor definitely rule out that the burning layer 
already shows significantly enhanced  helium. We suggest that this should be assessed
in the future with observations at other wavelengths, for instance
detecting the He II $\lambda$1640 absorption line in the ultraviolet  spectrum
when the white dwarf is cooling.

One other interesting question remains open.
What is the difference between a classical nova like N LMC 1995 and GQ Mus,
that burns hydrogen in a shell for several years after the outburst, 
and the majority
of novae, for which turn off occurs  within a couple of years? 
\"Ogelman et al. (1993) speculated that for GQ Mus hydrogen burning 
is rekindled because of irradiation induced mass transfer from the
companion. This mechanism, however, is efficient  for systems 
 with orbital periods
shorter than 4 hours (Kovetz et al. 1988), but it does not seem to operate
efficiently when  the orbital
period, like probably in this nova, is larger (Lipkin \& Orio 2003, in preparation). 
The theoretical models foresee that
the white dwarf mass is the main parameter that determines
the turn-off time and the amount of 
accreted mass retained after each outburst.
We hope to verify it, thus answering one of the most basic questions of nova physics,
 by monitoring the further evolution of N LMC 1995 in X-rays,
 and possibly at the same time in optical and ultraviolet. 
 
\acknowledgments

 M.O. is  grateful to Sumner Starrfield for a useful discussion.
This research has been supported by a NASA grant for {\sl XMM}
guest observations.

\clearpage


\clearpage 

\begin{figure}
\hbox{\hspace{2.cm}{\epsfig{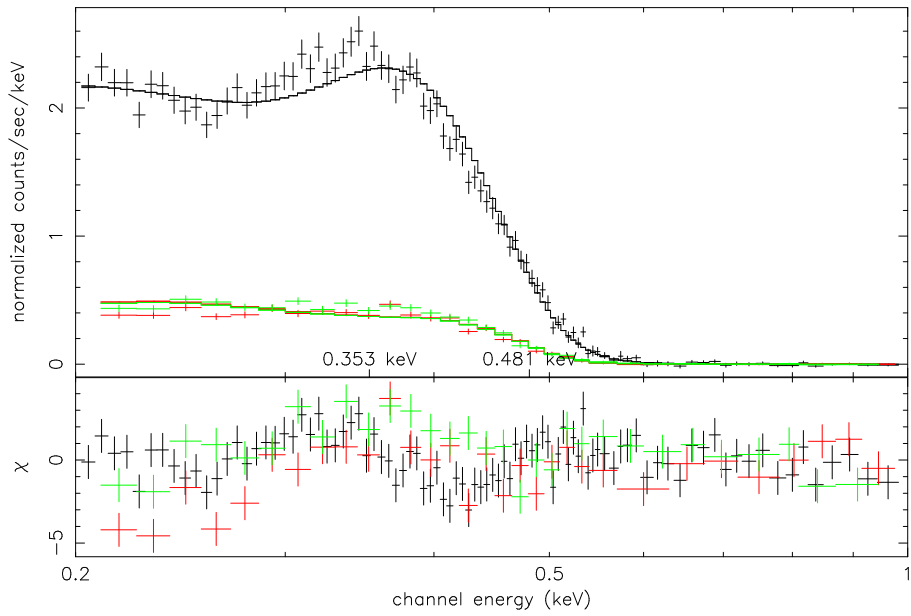}}}
\caption{ The background corrected count  rate spectra   of N LMC   1995
  observed in the 0.2-1 keV energy range with  {\sl XMM EPIC pn}
(in black)  and
{\sl EPIC MOS-1 and MOS-2}
(respectively, in red and green) on    December  19  2000.
We also show the fit with model
``b''  and log(g)=8 (obtained simultaneously fitting
 the {\sl RGS} data and excluding the range 0.2-0.3 keV for {\sl MOS}
data). In this NLTE atmosphere, we assume T$_{\rm  eff}$=407,000  K
 and  N(H)=9 $\times$ 10$^{20}$ cm$^{-2}$.
\label{fig1}}
\end{figure}

\clearpage

\begin{figure}[t]
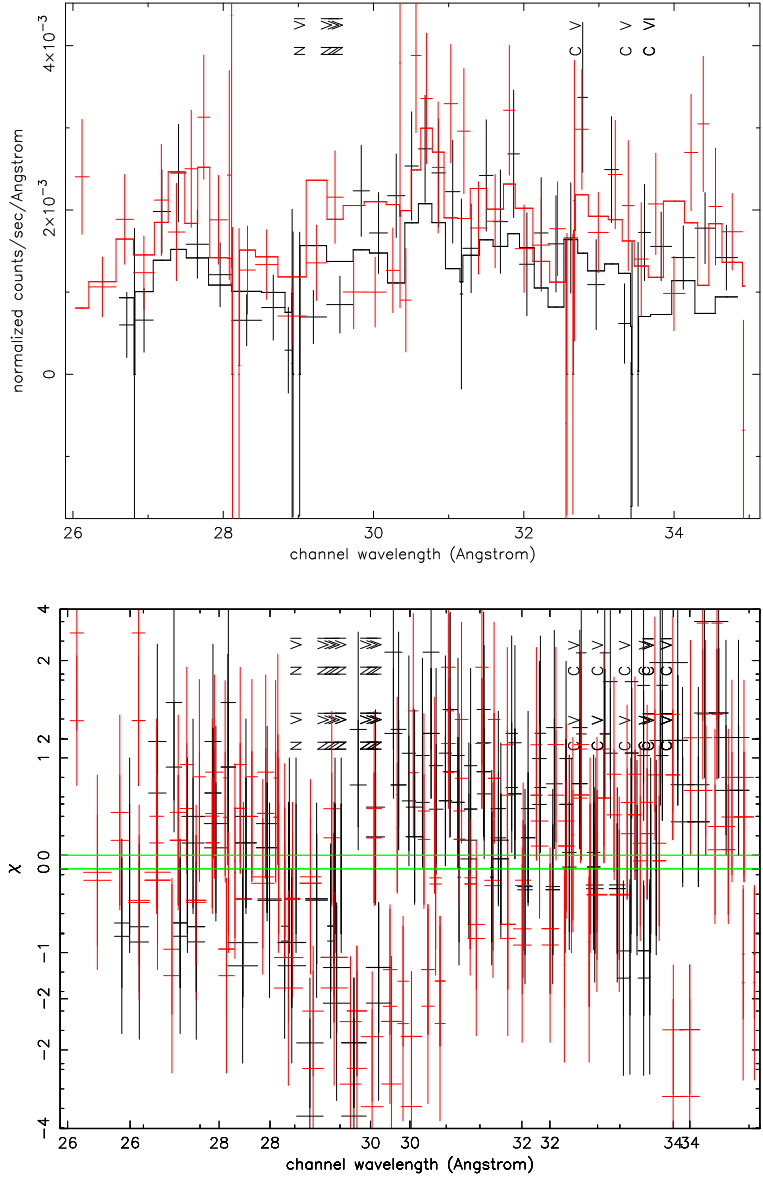

\hbox{\hspace{2.8cm}\epsfig{figure=f2a.eps,angle=-90,width=10cm}}
%
\vspace{0.5cm}\hspace{2.8cm}{\epsfig{figure=f2b.eps,angle=-90,width=10cm}}
\caption{ The panel on the top shows the background corrected
count  rate spectra   of N LMC   1995  observed with  {\sl XMM RGS-1} 
(in black) and with {\sl RGS-2} (in red) 
 on    December  19  2000 in the wavelength range 25-35 \AA 
(energy range 0.353-0.481 keV).
The best fit with model ``b'' (log(g)=8)
to the spectrum observed with
 {\it all} the instruments (see Fig 1 and Table 3) is superimposed
(folded with the response of both instruments). The panel on the bottom shows the
residuals with each instruments at each wavelength.
 The wavelengths at which some of the strongest lines due to H-like  and
He-like transitions would be detected assuming  a temperature 407,0000 K,
are indicated.  \label{fig2}}
\end{figure}

\clearpage

\begin{figure}
\hbox{\hspace{2.cm}{\epsfig{figure=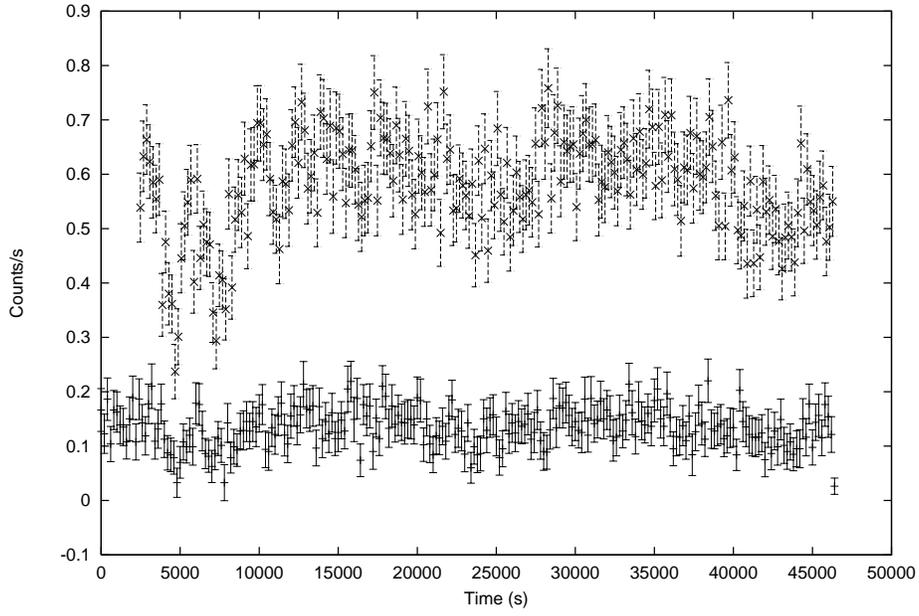,angle=-90,width=12cm}}}
\caption{The {\sl EPIC-pn} and {\sl EPIC-MOS-1} background corrected
light curves during the observation, binned every 200 s. 
  \label{fig3}}
\end{figure}

\clearpage

\begin{table}
\begin{center}
 \caption{Average background corrected count rates
measured with the {\sl Newton-XMM} instruments.\label{tbl-1}}
\begin{tabular}{crrr}
\tableline\tableline
 Instrument & Range (keV) & Count rate (cts s$^{-1}$) \\
\tableline
  EPIC pn & 0.2-10  & 0.5754$\pm$0.0043 \\
  EPIC MOS-1 & 0.2-10 & 0.1025$\pm$0.0017 \\
  EPIC MOS-2 & 0.2-10 & 0.1144$\pm$0.0017  \\
  RGS-1    & 0.3-3.5 & 0.0196$\pm$0.0010 \\
  RGS-2    & 0.3-3.5 & 0.0221$\pm$0.0010 \\
\tableline
 \end{tabular}
 \end{center}
\end{table}

\clearpage
 \begin{table}
\begin{center}
 \caption{ Species of ions, number of levels considered,
their main quantum number, number of lines used in
the preliminary  atmosphere calculation with
  temperature  400,000 K, number of lines used in the spectrum
calculation between 20 and 40 \AA \ (see Hartmann 
et al. 1999 for details). Note that the exact number of lines
in the spectrum varies with T$_{\rm eff}$. \label{tbl-2}}
\begin{tabular}{crrrrr}
\tableline\tableline
 Ion & Levels number & Quantum number & Lines (atmosphere) & Lines (spectrum) \\
\tableline
  HI     & 9   &  1-9       &   28   &     0 \\
  HeII   & 14  &  1-14      &   78   &     0 \\
  CV     & 5   &  1-2       &    3   &     8 \\
  CVI    & 6   &  1-3       &    5   &     9 \\ 
  NV     & 5   &  2-3       &    6   &     0 \\
  NVI    & 5   &  1-2       &    3   &     9 \\
  NVII    & 6 & 1-3       &    5   &     2 \\
  OVI    &  5 & 2-3       &    6   &     0 \\
  OVII   &  5 & 1-2       &    3   &     1 \\
  NeVIII &  5 & 2-3       &    6   &     0 \\
   NeIX  &  5 & 1-2       &    3   &     0 \\
   MgX   &  5 & 2-3       &    6   &    48 \\
  Mg XI  &  5 & 1-2       & 3        & 40 \\ 
SX    &     8   &  2      &    9  &   429  \\
SXI   &     12  &  2      &   12  &    611 \\
SXII  &      8  &  2      &    9  &    391 \\
SXIII &   6     &  2      &    4  &    158 \\
ArX   &   2     &  2      &    1  &     91 \\
ArXI  &   6     &  2      &    4  &    353 \\
ArX   &   2     &  2      &    1  &     91 \\
ArXI  &  6      &  2      &    4  &    353 \\
ArXII &  8      &  2      &    9  &    676 \\
CaX   &  7      & 3-4     &   11  &      0 \\
CaXI  & 15      & 2-3     &   20  &     18 \\
CaXII & 12      & 2       &    1  &     84 \\
FeXV  & 14      & 3       &  19   &    206 \\
FeXVI &  7      & 3-4     &  11   &     24 \\
FeXVII &  15    & 2-3     &  20   &    382 \\
Fe XIV & 6      & 3       &  4    &    284 \\
\tableline   
 \end{tabular}
 \end{center}
\end{table}

\clearpage
 
\begin{table}
\begin{center}
\caption{
Parameters of the NLTE atmospheric models
(described in the text) fitted to the observed spectra,
for models which yield  $\chi^2$/d.o.f.$\leq$1.5 at least
in the 0.3-1.0 spectral range. 
Best fit temperature and column
of neutral hydrogen N(H) (given with the 2$\sigma$ error
unless N(H) if it was ``frozen'', in which case it is marked 
with an asterisk), and normalization constant K=R$_{\rm WD}^2$/d$^2$.
The flux was multiplied by  a constant, normalized to 
1 for the {\sl pn}, and let vary as  a free parameter 
for the other instruments. It turned out to be about 1.05 for {\sl MOS-1},
1.15 for {\sl MOS-2}, 0.65 for both two {\sl RGS},  respectively. 
The spectrum was always fitted above 0.3 keV for the two {\sl MOS}, in
the range 0.35-0.48 keV for the two {\sl RGS}, 
and if the {\sl pn} is included in the
 fit, the ``range'' reported  in column 2 is for the {\sl pn}.
\label{tbl-3}}
\begin{tabular}{crrrrrrrr}
\tableline\tableline
 Instrument & range & Model & log(g) & T$_{\rm eff} \times 10^3$ & N(H) 
$\times 10^{21}$ &  K $\times 10^{-29}$  & $\chi^2$/d.o.f.\\ 
    & (keV)    &       &     &  (K)  & (cm$^{-2}$)   & & \\ 
\tableline
pn  &             0.3-1.0   & b   & 8.0  & 411$\pm$1  & 0.70$\pm$0.04 & 
2.59  & 1.38 \\
pn  &             0.3-1.0   & d   & 8.0  & 402$\pm$1  & 1.80* & 341.99 
   & 1.07 \\
pn  &             0.3-1.0   & d   & 8.5  & 452$\pm$2  & 1.09$\pm$0.13 & 
 24.97   & 1.10 \\
pn  &             0.3-1.0   & d   & 9.0  & 454$^{+8}_{-7}$ & 1.48$\pm$0.13
  & 70.86   & 0.99 \\
pn  &             0.3-1.0   & d   & 8.5  & 454$\pm$2  & 1.09$\pm$0.13 & 
 24.97  & 1.10 \\
pn  &             0.2-1.0   & d   & 8.5  & 452$^{+1}_{-6}$
  & 1.41$^{+0.06}_{-0.04}$ & 48.54    & 1.33 \\ 
pn  &             0.2-1.0   & d   & 9.0  & 452$^{+2}_{-5}$  & 
1.57$^{+0.08}_{-0.06}$ & 91.33    & 1.09 \\
MOS &        0.3-1.0   & a   & 9.0  & 417$\pm6$  & 1.80* & 321.66    & 1.29 \\
MOS &        0.3-1.0   & b   & 9.0  & 399$^{+1}_{-14}$
  & 1.80* &  57.19   & 1.26 \\
MOS &        0.3-1.0   & d   & 8.5  & 446$^{+5}_{-14}$  & 
1.48$^{+0.08}_{-0.07}$ & 84.28  & 1.40 \\
pn + MOS &   0.3-1.0 & b & 8.0 & 411$\pm$2  & 0.70$\pm$0.05 
& 2.84 & 1.07 \\
pn + MOS &   0.2-1.0 & d & 8.5 & 451$\pm 1$ & 1.45$^{+0.04}_{-0.05}$  
& 54.45   & 1.40 \\
pn + MOS &   0.2-1.0 & d & 9.0 & 451$^{+1}_{-4}$ & 1.64${\pm 0.07}$
 & 113.45   & 1.27 \\
RGS &        0.35-0.48    & b & 8 & 410$\pm$6       & 0.70*                  &
 2.23      & 1.28 \\
RGS &        0.35-0.48    & d & 8 & 402$^{+9}_{-7}$ & 1.34$^{+0.15}_{-0.50}$ &
 165.19  & 1.41 \\
RGS &        0.35-0.48    & d & 8.5 & 454$^{+9}_{-4}$ & 0.75$\pm 0.12$ & 
 10.83  & 1.40 \\
RGS &        0.35-0.48    & d & 9.0 & 477$^{+23}_{-18}$ & 0.69$\pm 0.34$ 
  & 6.24  & 1.25 \\     
RGS &        0.35-0.48    & e & 8.0 & 416$^{+7}_{-5}$ & 0.70* & 0.75 & 1.28 \\
RGS &        0.35-0.48    & e & 8.5 & 460$^{+38}_{-4}$ & 
0.74$^{+0.13}_{-0.05}$ & 0.40  & 1.33 \\
RGS &        0.35-0.48    & e & 9.0 & 473$\pm15$ & 
1.18$\pm$0.35 & 0.74  & 1.37 \\
 pn + RGS &  0.3-1.0       & b & 8.0 & 411$\pm2$        & 0.70* & 2.58
 & 1.17 \\
 pn + RGS &  0.2-1.0       & b & 8.0 & 407$\pm2$        & 0.90$\pm$0.02 & 4.87 
 & 1.53 \\ 
 pn + RGS &  0.2-1.0       & d & 8.5 & 452$^{+1}_{-7}$  & 1.41$\pm0.05$ &
 49.60 & 1.38 \\
 pn + RGS &  0.2-1.0      & d & 9.0 & 453$\pm$2       & 1.54$^{+0.06}_{-0.02}$ &
 83.82  & 1.27 \\
ALL &  0.3-1.0             & b & 8.0 & 409$\pm$1       & 0.70*         &
 2.75  & 1.30 \\ 
ALL &  0.2-1.0             & d & 8.5 & 451$^{+2}_{-7}$ & 1.46$\pm$0.05 &
 53.9  & 1.42 \\ 
ALL &  0.2-1.0             & b & 8.0 & 407$\pm$2       & 0.90$\pm$0.02 & 4.87 
& 1.53 \\
ALL &  0.2-1.0             & d & 8.5 & 451$^{+1}_{-7}$ & 1.45$\pm$0.05 &
 54.48 & 1.41 \\
ALL &  0.2-1.0             & d & 9.0 & 451$\pm$1 & 1.60$\pm$0.08 & 
102.39 & 1.35 \\
\tableline
\end{tabular}
\end{center}
\end{table}

\end{document}